# Quantum diffusion

Roumen Tsekov
Department of Physical Chemistry, University of Sofia, 1164 Sofia, Bulgaria

Quantum diffusion is studied via dissipative Madelung hydrodynamics. Initially the wave packet spreads ballistically, than passes for an instant through normal diffusion and later tends asymptotically to a sub-diffusive law. It is shown that the apparent quantum diffusion coefficient is not a universal physical parameter since it depends on the initial wave packet preparation. The overdamped quantum diffusion of an electron in the field of a periodic potential is also investigated; in this case the wave packet spreads logarithmically in time. Thermo-quantum diffusion of heavier particles as hydrogen, deuterium and tritium atoms in periodic potentials is studied and a simple estimate of the tunneling effect is obtained in the frames of a quasi-equilibrium semiclassical approach. The effective thermo-quantum temperature is also discussed in relation to the known temperature dependence of muon diffusivity in solids.

Quantum diffusion (QD) describes a wave packet spreading in a dissipative environment at zero temperature. Since quantum effects are significant for light particles mainly, QD is very essential for electrons, which on the other hand are very important in physics and chemistry. QD has been experimentally observed, however, for muons as well [1, 2], which are about 200 times heavier than electrons. Studies on electron transport in solids are strongly motivated by the semiconductor industry, exploring nowadays quantum effects on nano-scale. A contemporary review on electron quantum diffusion in semiconductors is given in [3]. Traditionally, QD is theoretically described by means of the models of quantum state diffusion [4], quantum Brownian motion [5], quantum drift-diffusion [6], etc.

Another important transport process affected by quantum effects is the diffusion of hydrogen atoms or molecules in metals and on solid surfaces [7, 8]. The quantum tunneling accelerates the hydrogen diffusion, which is essential for many modern technologies for storage and use of hydrogen as a fuel, chemical reagent, etc. An extensive collection of papers on surface diffusion of atoms is given in [9], where hydrogen takes naturally a central place. Since the atoms are much heavier than electrons their diffusion can be described well by semiclassical considerations [10-14]. The interplay of thermal activation and quantum tunneling effects results in thermo-quantum diffusion, which plays important role not only for transport but also in chemical kinetics, shape-selective catalysis, crystal growth, isotope separation, etc.

The scope of the present paper is to explore a new way for QD description based on the quantum hydrodynamics. The latter has been proposed first by Madelung [15] and in the recent years it has grown to a modern quantum theoretical approach [4, 16]. The quantum hydrodynamics is very similar to the de Broglie-Bohm theory, which nowadays is considered by many physicists as a challenge to the conventional Bohr-Heisenberg quantum mechanics. Our basic contribution to quantum hydrodynamics here consists in inclusion of dissipative and thermal forces [17, 18]. Thus, important novel results are obtained either for quantum diffusion, typical for electrons, or for thermo-quantum diffusion, typical for hydrogen atoms. The comparison of the latter with some experimental observations confirms the correctness of the present theoretical model, which on the other hand represents an alternative of the existing semiclassical models. Finally, the effective thermo-quantum temperature is discussed in relation to the reported in literature temperature dependence of diffusivity of muons in solids. This concept is already applied in descriptions of quantum Brownian motion [19] and chemical kinetics [20].

## Quantum diffusion

If a quantum particle, an electron for instance, moves in vacuum its wave function $\psi$ evolves according to the Schrödinger equation

$$i\hbar \partial_t \psi = (-\hbar^2 \nabla^2 / 2m + U)\psi \tag{1}$$

where $m$ is the particle mass and $U$ is an external potential. Since the wave function is complex it can be generally presented in the polar form $\psi(r,t) = \sqrt{\rho} \exp(iS/\hbar)$ [15], where $\rho(r,t)$ is the probability density to find the quantum particle in a given point $r$ at time $t$ and $S(r,t)$ is the wave function phase. Introducing this presentation in Eq. (1) results rigorously in the following two equations [15], corresponding to the imaginary and real parts, respectively,

$$\partial_t \rho + \nabla \cdot (\rho V) = 0 \tag{2}$$

$$m\partial_t V + mV \cdot \nabla V = -\nabla U - \nabla \cdot \mathbb{P}_Q / \rho \tag{3}$$

Equation (2) is a continuity equation. Therefore, the velocity $V \equiv \nabla S / m$ represents the flow in the probability space and Eq. (3) is its hydrodynamic-like force balance. As is seen, the quantum effect is completely included in the quantum pressure tensor $\mathbb{P}_Q \equiv -(\hbar^2 / 4m)\rho \nabla \otimes \nabla \ln \rho$. The description of the quantum evolution by these Madelung equations is identical to the Schrödinger picture and it is a matter of convenience which method will be employed. The Madelung hydrodynamics differs [21], however, from the de Broglie-Bohm theory [22], where $V$ is associated to the real velocity of a quantum particle.

Equation (3) describes the force balance in vacuum. If the quantum particle moves in a dissipative environment it will experience also a friction force, which is generally proportional to the particle velocity. Hence, the corresponding generalization of Eq. (3) reads [23]

$$m\partial_t V + mV \cdot \nabla V + bV = -\nabla(U+Q) \qquad (4)$$

where $b$ is the particle friction constant and $Q \equiv -\hbar^2 \nabla^2 \sqrt{\rho}/2m\sqrt{\rho}$ is the Bohm quantum potential [22] being related to the quantum pressure tensor via the relation $\nabla \cdot \mathbb{P}_Q = \rho \nabla Q$. Thus, the system of Eqs. (2) and (4) describes the probability spreading in a dissipative environment (QD). A speculative reversal back to a wave function via the Madelung presentation leads to a nonlinear Schrödinger equation [23]. In the case of a free quantum particle ($U=0$) the probability density is Gaussian $\rho = \exp(-r^2/2\sigma^2)/(2\pi\sigma^2)^{3/2}$, where $\sigma^2(t)$ is the dispersion of the wave packet. Substitution of this expression in Eq. (2) results in an expression for the hydrodynamic-like velocity $V = r\partial_t \ln \sigma$. Introducing now both expressions for $\rho$ and $V$ in Eq. (4) yields the following equation

$$m\partial_t^2 \sigma + b\partial_t \sigma = \hbar^2/4m\sigma^3 \qquad (5)$$

describing the evolution of the root-mean-square displacement $\sigma$. Equation (5) is already derived before [24-26, 17]. Introducing new dimensionless dispersion $\xi^2 \equiv 2b\sigma^2/\hbar$ and time $\tau \equiv bt/m$ Eq. (5) acquires the universal form of a dissipative Ermakov equation

$$\partial_\tau^2 \xi + \partial_\tau \xi = \xi^{-3} \qquad (6)$$

The initial conditions relevant to Eq. (6) are $(\partial_\tau \xi)_0 = 0$ and $\xi_0 = \sqrt{2b/\hbar}\,\sigma_0$.

It seems that Eq. (6) cannot be solved analytically and for this reason let us consider first some limiting cases. If $\tau \ll 1$ than $\partial_\tau^2 \xi \gg \partial_\tau \xi$ and Eq. (6) reduces to $\partial_\tau^2 \xi = \xi^{-3}$. The solution of this equation $\xi^2 = \xi_0^2 + \tau^2/\xi_0^2$ corresponds to the well-known expression $\sigma^2 = \sigma_0^2 + (\hbar t/2m\sigma_0)^2$ for spreading of a Gaussian wave packet in vacuum. In the opposite case at large times $\tau \gg 1$ one can neglect the acceleration $\partial_\tau^2 \xi \ll \partial_\tau \xi$ and Eq. (6) reduces to $\partial_\tau \xi = \xi^{-3}$. The solution of this equation $\xi^4 = \xi_0^4 + 4\tau$ corresponds to the expression $\sigma^2 = \sqrt{\sigma_0^4 + \hbar^2 t/mb}$. Hence, at short time the dispersion evolves as classical diffusion $\sigma^2 = \sigma_0^2 + \hbar^2 t/2mb\sigma_0^2$ with a quantum diffusion constant, while at large time it tends to the sub-diffusive law $\sigma^2 = \hbar\sqrt{t/mb}$ for the overdamped quantum diffusion [17]. In Fig. 1 the numerical solution of Eq. (6) at $\xi_0^2 = 0.1$ is plotted in the form $\xi^2$ vs. $\tau$ as well as the time-derivative $\partial_\tau \xi^2$. As is seen initially the dispersion $\xi^2$ increases ballistically in time as in vacuum, later the plot is almost linear and at the end at large times the dispersion increases according to the asymptotic law $\xi^2 = 2\sqrt{\tau}$. The maximum of the derivative $\partial_\tau \xi^2$ corresponds to something that is called quantum diffusion constant since at $\partial_\tau^2 \xi^2 = 0$ $\xi^2$ increases linearly with $\tau$.

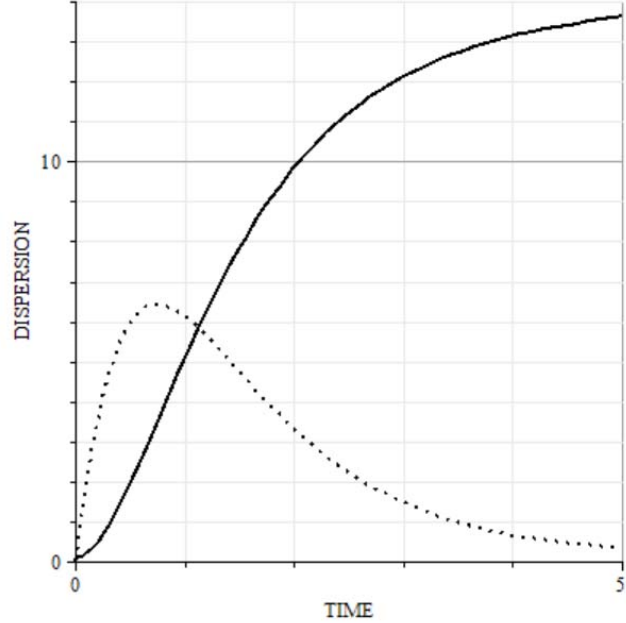

**Fig. 1** Evolution of the dimensionless dispersion $\xi^2$ (solid line) and its rate of change $\partial_\tau \xi^2$ (dotted line) on the dimensionless time $\tau$ for initial dispersion $\xi_0^2 = 0.1$.

In Fig. 2 the dependence of $(\partial_\tau \xi^2)_{max}$ vs. $\xi_0^2$ is plotted as calculated numerically from Eq. (6). It is seen that the quantum diffusion constant decreases with increasing initial dispersion $\xi_0^2$ and a good fit of this dependence is $(\partial_\tau \xi^2)_{max} = 1/2\xi_0^2$ for $\xi_0^2 \leq 0.1$. Hence, the apparent quantum diffusion constant equals to

$$D_Q \equiv (\partial_t \sigma^2)_{max}/2 = \hbar^2/16mb\sigma_0^2 \qquad (7)$$

This expression shows that, in contrast to the classical Einstein diffusion constant $D = k_B T/b$, the quantum diffusion constant $D_Q$ is not a universal physical parameter and depends on the initial preparation of the wave packet via the dispersion $\sigma_0^2$. This could explain, for instance, the large spreading of quantum surface diffusion coefficients measured at low temperatures [27].

Perhaps different experimental techniques correspond to different initial preparations and the variations of the initial dispersion $\sigma_0^2$ result in different values of $D_Q$. The ratio between the quantum and Einstein diffusion coefficients $D_Q/D = (\lambda_T/2\sigma_0)^2$ scales with the ratio between square of the thermal de Broglie wave length $\lambda_T = \hbar/2\sqrt{mk_BT}$ and the initial dispersion $\sigma_0^2$.

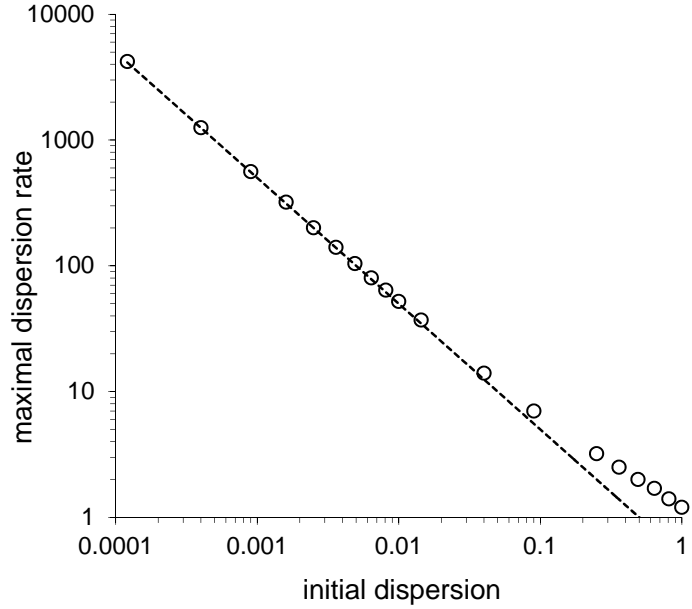

**Fig. 2** Dependence of the maximal dimensionless rate $(\partial_\tau \xi^2)_{max}$ on the initial dimensionless dispersion $\xi_0^2$. The line corresponds to $(\partial_\tau \xi^2)_{max} = 1/2\xi_0^2$.

Another interesting case is a quantum harmonic oscillator with potential $U = m\omega_0^2 r^2/2$. In this case the probability density is Gaussian again and the analog of Eq. (6) reads [18, 24-26]

$$\partial_\tau^2 \xi + \partial_\tau \xi + \alpha^2 \xi = \xi^{-3} \tag{8}$$

where $\alpha \equiv m\omega_0/b$ is a dimensionless parameter dependent on the oscillator own frequency $\omega_0$ and the momentum relaxation time $m/b$. This equation is known in the literature as damped Pinney equation [28]. Obviously, the equilibrium solution of Eq. (8) at $\tau \to \infty$ is $\xi_\infty^2 = \alpha^{-1}$, which corresponds to the position dispersion $\sigma_\infty^2 = \hbar/2m\omega_0$ in the oscillator ground state. Hence, due to the friction force the oscillator loses its energy until it drops at the zero point level. This is evident from Fig. 3, where a numerical solution of Eq. (8) is presented for the particular case $\alpha = 1$. Indeed, after several damped oscillations the oscillator reaches its ground state $\xi_\infty^2 = 1$.

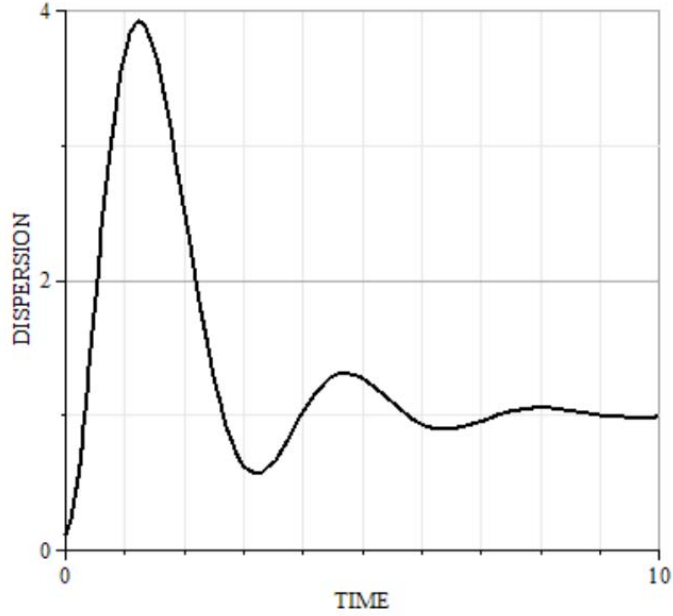

**Fig. 3** Evolution of the dimensionless dispersion $\xi^2$ of a harmonic oscillator on the dimensionless time $\tau$ for initial dispersion $\xi_0^2 = 0.1$ and $\alpha = 1$.

As was mentioned before the usual quantum diffusion goes under the action of a periodic potential $U$. It is difficult to solve in this case Eqs. (2) and (4). Since usually the friction force is strong one can neglect the first two inertial terms in Eq. (4) and introducing now the obtained expression for the velocity $V$ in Eq. (2) yields a quantum diffusion equation

$$\partial_t \rho = \nabla \cdot [\rho \nabla (U + Q)/b] = \nabla \cdot (\rho \nabla U + \nabla \cdot \mathbb{P}_Q)/b \tag{9}$$

Due to the quantum terms Eq. (9) is a nonlinear equation, which is still difficult to solve. To linearize it one can use the fact that the effective shape of the probability density is expected to be Gaussian in the case of relatively small potentials $U$, which corresponds to lack of localization. Hence, one can employ the approximation $\nabla \otimes \nabla \ln \rho = -\mathbb{I}/\sigma^2$, where $\sigma^2$ is the dispersion of the probability density and $\mathbb{I}$ is the unit tensor. Introducing this expression in the quantum pressure tensor $\mathbb{P}_Q$ in Eq. (9) yields a Smoluchowski equation

$$\partial_t \rho = \nabla \cdot (\rho \nabla U + \beta_Q^{-1} \nabla \rho)/b \tag{10}$$

with a quantum thermodynamic-like temperature $\beta_Q^{-1} \equiv \hbar^2 / 4m\sigma^2$ corresponding to the momentum dispersion from the minimal Heisenberg relation valid for Gaussian processes. Lifson and Jackson [29] have derived a general formula for calculation of the effective diffusion coeffi-cient from the Smoluchowski equation for periodic potentials. Since $\beta_Q$ depends on time via $\sigma^2$ one should use a more general expression derived by Festa and d'Agliano [30]

$$\partial_t \sigma^2 / 2 = [b\beta_Q <\exp(\beta_Q U)><\exp(-\beta_Q U)>]^{-1} = [b\beta_Q I_0^2(\beta_Q A)]^{-1} \tag{11}$$

where the brackets $<\cdot>$ indicate a spatial geometric average. The last expression is obtained particularly for the cosine potential $U = A\cos(qx)$ and $I_0$ is the modified Bessel function of first kind and zero order. An integration of Eq. (11) on time yields the relation

$$(\beta_Q A)^2 [I_0^2(\beta_Q A) - I_1^2(\beta_Q A)] = 16mA^2 t / \hbar^2 b \tag{12}$$

where $I_1$ is the modified Bessel function of first kind and first order.

Equation (12) reduces to the already mentioned expression $\sigma^2 = \hbar\sqrt{t/mb}$ for the purely quantum diffusion in a non-structured environment [17] at $A = 0$. In the opposite case $\beta_Q A > 1$ the modified Bessel functions difference is well approximated by $I_0^2(x) - I_1^2(x) \approx \exp(2x)/2\pi x^2$. Thus according to Eq. (12) the dispersion of the purely quantum diffusion in a strong cosine potential depends logarithmically on time

$$\sigma^2 = (\hbar^2 / 8mA)\ln(32\pi mA^2 t / \hbar^2 b) \tag{13}$$

This result demonstrates a lack of classical diffusive behavior. As is seen from Eq. (13), the magnitude of the deviation $\sigma$ scales with the de Broglie wave length $\lambda_A = \hbar/2\sqrt{2mA}$ of the activation energy [31], while its relaxation time $b/m\omega_A^2$ corresponds to that of a harmonic oscillator with an own frequency $\omega_A = 4A/\hbar$. Even if this logarithmic dependence is derived here particularly for a cosine external potential we believe that it is general for any periodic potential and reflects the Arrhenius law. Indeed the latter holds in any periodic potential when the potential barriers are much higher than the thermal energy and the exact type of the potential affects the pre-exponential factor only [29, 30]. Hence, QD does not obey the classical linear Einstein law of Brownian motion in periodic potentials as well.

### Thermo-quantum diffusion

How it was shown above the diffusion in the field of periodic potentials is important for particle transport via ordered structures, typical in condensed matter, and the underlying mechanism is Brownian motion. If the diffusing particles are quantum ones, e.g. electrons, protons, atoms, etc., their migration is described via the theory of quantum Brownian motion [11-

14, 32-35], which is significantly affected by the tunneling effect. Hereafter an alternative approach to the problem of thermo-quantum Brownian motion in periodic potentials is developed based on the Bohm quantum potential. In contrast to the WKB theory valid in vacuum the present paper describes tunneling in a dissipative environment. The starting point is a nonlinear quantum Smoluchowski-like equation [18, 34]

$$\partial_t \rho = D\nabla \cdot [\rho \nabla \int_0^\beta (U+Q)_b \, d\beta + \nabla \rho] \tag{14}$$

being a generalization of Eq. (9) for finite temperatures. Here $\beta = 1/k_B T$ and $D = k_B T / b$ is the Einstein diffusion constant. The goal of the present analysis is to estimate the diffusion coefficient from Eq. (14) for the case of the periodic cosine potential $U = A\cos(qx)$. Since due to the quantum potential Eq. (14) is nonlinear, obtaining its general solution is mathematically frustrated. In a semiclassical approach one could estimate $Q$ by using the classical solution of Eq. (14). However, even in the classical case it is impossible to obtain a simple physically transparent expression for the probability density. For this reason, we are going to apply a stronger approximation by employing the equilibrium classical Boltzmann distribution $\rho_{cl}^{eq} \propto \exp(-\beta U)$ into the quantum potential to obtain

$$Q(\rho_{cl}^{eq}) = \lambda_T^2 [\nabla^2 U - \beta (\nabla U)^2 / 2] = -\lambda_T^2 q^2 [U + \beta(A^2 - U^2)/2] \tag{15}$$

The first expression here is general, while the last one is valid for the present particular cosine potential. Substituting Eq. (15) in Eq. (14) and performing the integration on $\beta$ results in the following quasi-equilibrium semiclassical Smoluchowski equation

$$\partial_t \rho = \nabla \cdot (\rho \nabla U_{eff} + \beta^{-1} \nabla \rho)/b \tag{16}$$

where the new effective potential is given by $U_{eff} = [1 - \lambda_T^2 q^2 (1 - \beta U / 3)/2]U$ and possesses the same periodicity as the external potential $U$.

How it was mentioned in the previous section, there is a general formula for calculation of the diffusion coefficient from Eq. (16) applied for periodic potentials [29, 30]

$$D_{eff} = D / [<\exp(\beta U_{eff}) > < \exp(-\beta U_{eff}) >] \tag{17}$$

At room temperature one can neglect the small nonlinear contribution of the external potential to the effective potential and thus the latter acquires a simpler form $U_{eff} = (1-\lambda_T^2 q^2/2)U$. In this case the spatial averaging in Eq. (17) can be analytically accomplished and the result reads

$$D_{eff} = D / I_0^2[\beta A(1-\lambda_T^2 q^2/2)] \qquad (18)$$

In the case of free Brownian motion ($A = 0$) Eq. (18) provides the classical Einstein constant $D$, which is due to the quasi-equilibrium semiclassical approach [34]. At large arguments the modified Bessel function can be approximated by $I_0(x) \approx \exp(x)/\sqrt{2\pi x}$ and Eq. (18) reduces in this case to the Arrhenius law

$$D_{eff} = \pi(2-\lambda_T^2 q^2)(A/b)\exp[-\beta(2-\lambda_T^2 q^2)A] \qquad (19)$$

As is seen from Eq. (19), both the activation energy $E_a = (2-\lambda_T^2 q^2)A$ and pre-exponential factor $\pi(2-\lambda_T^2 q^2)A/b$ are affected by quantum effects. In the classical limit $\lambda_T \to 0$ and the activation energy $E_a = 2A$ equals to the difference between the maximal and minimal value of the external potential. The quantum effect decreases effectively the activation energy due to quantum tunneling. For instance, the thermal de Broglie wave length for a proton at room temperature equals to $\lambda_T \approx 0.2$ Å. If protons are diffusing in a structured medium with a lattice constant 3.6 Å then $\lambda_T^2 q^2 \approx 0.1$. Hence, the tunneling effect will decrease the activation energy by 5 % as well as the pre-exponential factor. Note that if $\lambda_T^2 q^2 = 2$ the diffusion is free since $U_{eff} = 0$ which for a proton corresponds to $T \approx 18$ K.

Let us consider the experimental result for diffusion of hydrogen and deuterium on a Ni (111) surface [8]. Using the diffusion coefficients experimentally measured at high temperature one can estimate from the classical activation energy $E_a = 20$ kJ/mol the potential parameter $A = 1.67 \times 10^{-20}$ J, while from the pre-exponential factor $D_0 = 3.2 \times 10^{-7}$ m²/s the friction coefficient $b = 3.3 \times 10^{-13}$ kg/s can be estimated. Thus the characteristic momentum relaxation time of a hydrogen atom, for instance, can be calculated as $m/b = 5 \times 10^{-15}$ s. If we use the reasonable value 3.6 Å for the lattice constant one can calculate the effective diffusion coefficient from Eq. (19). Its temperature dependence is plotted on Fig. 4 in Arrhenius coordinates for hydrogen, deuterium and tritium atoms. As seen the mass of the atom decreases substantially the tunneling effect. The diffusion constants correspond well to the experimentally measured values [8].

According to Eq. (19) the effective activation energy $E_a = (2-\lambda_T^2 q^2)A$ changes continuously with temperature and hence there is no quantum activation energy as usually considered [8]. Similar dependence follows from the semiclassical theory [35]. For electrons the thermal de

Broglie wave length at room temperature is $\lambda_T \approx 8.6$ Å and, hence, $\lambda_T^2 q^2 \approx 225$. Since this number is much larger than 2, the conclusion is that the present thermo-quantum theory is not applicable to electrons since they are essentially quantum particles and cannot be described by a quasi-equilibrium semiclassical approach. An alternative interpretation of the tunneling effect from Eq. (19) goes via the effective reciprocal temperature $\beta_{eff} = \beta(1 - \lambda_T^2 q^2 / 2)$ which possesses a maximum in respect to the standard temperature at $T_q = \hbar^2 q^2 / 4mk_B$. A related minimum in the diffusivity is experimentally detected for muons in Cu at temperature about 80 K [1, 2], which corresponds according to the above formula to a potential period of $7.2$ Å. This means that the potential barriers for muons come at every second Cu atom. As is seen from Fig. 4, for protons $T_q$ is about 37 K. However, the predictions of Eq. (19) for temperatures below $T_q$ are not very accurate since quantum effects dominate and could not be described anymore via a quasi-equilibrium semiclassical approach. Indeed, the experimental studies of hydrogen diffusion on W show that at lower temperatures the thermal diffusivity is negligible and the diffusion coefficient becomes almost temperature independent [7].

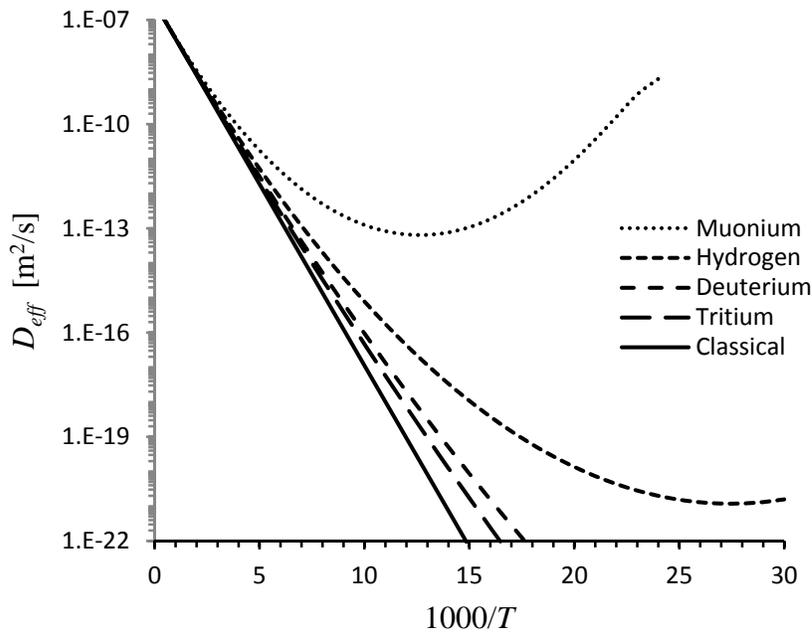

**Fig. 4** Temperature dependence of the effective semiclassical diffusion constant calculated from Eq. (19).

1. Lagos, M. and Rogan, J., Solid State Comm. **94**, 173 (1995)
2. Kadono, R., Appl. Magn. Reson. **13**, 37 (1997)
3. Kleinert, P., Phys. Rep. **485**, 1 (2010)